\documentstyle[prl,aps,graphicx]{revtex}

\begin{document}
\draft
\preprint{\vbox{\hbox{\bf NSC-NCTS-010219} }}

\title{Muon Anomalous Magnetic Moment and Leptoquark Solutions}
\author{Kingman Cheung}
\address{ National Center for Theoretical Science, National Tsing Hua
University, Hsinchu, Taiwan R.O.C.}
\maketitle

\begin{abstract}
The recent measurement on the muon anomalous magnetic moment $a_\mu$ shows a
$2.6\sigma$ deviation from the standard model value.
We show that it puts an interesting bound on the mass of the second generation
leptoquarks.  To account for the data the leptoquark must have both
the left- and right-handed couplings to the muon.
Assuming that the couplings have electromagnetic strength, the
mass is restricted in the range $0.7 \; {\rm TeV} < M_{\rm LQ} < 2.2$ TeV 
at 95\% C.L.
We also discuss constraints coming from other low energy and high 
energy experiments.  If the first-second-generation universality is assumed, 
constraints come from the atomic parity violation and
charged-current universality.  We show that coexistence with other leptoquarks
can satisfy these additional constraints and at the same time do not
affect the $a_\mu$.
\end{abstract}


Many Grand-Unified theories predict the existence of leptoquarks, which are
composite objects that carry both the lepton and quark numbers.  The discovery
of such particles certainly affects the planning for
future experiments and guides the building of the theories.  In fact, 
leptoquarks have been actively searched for in many collider experiments
\cite{teva-lq,hera-lq}, and will still be in the future.  
Precision measurements are also very useful in testing 
leptoquark models and restricting the parameter space.
The measurement of the anomalous magnetic moment of leptons \cite{lq2,equ}
is one of such experiments that can constrain the model.

The recent measurement on the muon anomalous
magnetic moment by the experiment E821 \cite{E821}
at Brookhaven National Laboratory 
has reduced the error to a substantially smaller level.  
Combining with previous measurements the new world average is \cite{am}
\begin{equation}
a_\mu^{\rm exp} = 116\, 592 \, 023\,(151) \; \times \; 10^{-11} \;,
\end{equation}
where the standard model (SM) prediction is 
\begin{equation}
a_\mu^{\rm SM} = 116\, 591 \, 597\,(67) \; \times \; 10^{-11} \;,
\end{equation}
in which the QED, hadronic, and electroweak contributions have been included.
Thus, the deviation from the SM value is
\begin{equation}
\Delta a_\mu \equiv 
a_\mu^{\rm exp}- a_\mu^{\rm SM} = 
(42.6\, \pm 16.5) \; \times \; 10^{-10} \;.
\end{equation}
This $2.6\sigma$ deviation may be a hint to new physics because the deviation
is beyond the uncertainties in QED, electroweak, and hadronic contributions.

Among various extensions of the SM, namely, supersymmetry \cite{susy},
additional gauge bosons \cite{Z}, leptoquarks \cite{lq2,ccc,lq}, 
extra dimensions, muon substructure \cite{preon},
they all contribute to $a_\mu$.
However, not all of them can contribute in the right direction as indicated
by the data.  Thus, the $a_\mu^{\rm exp}$ measurement can differentiate 
among various models, and perhaps with other existing data  can put very strong
constraints on the model under consideration.

In this Letter, we investigate the contributions of various leptoquarks to
$a_\mu$.  We limit to the second generation leptoquarks only without 
considering any generation mixing in order to avoid dangerous flavor changing
neutral currents.  
Our main result is summarized as follows.  To account for the $a_\mu$ data
the solution requires a leptoquark that has both the left-handed
and right-handed chiral couplings and the mass is required to be about 
 0.7 -- 2.2
TeV for an electromagnetic coupling strength.  This solution is consistent
with direct and indirect experimental search.  
The $a_\mu$ data disfavors, if not rule out, the leptoquarks that have only
a left- or right-handed coupling.  Also, 
coexistence with other leptoquarks can easily satisfy additional
constraints, e.g., atomic-parity violation (APV) and charged-current (CC)
universality, without affecting the $a_\mu$.

While we are completing this work, a paper \cite{ccc}
appears, which describes similar solutions to $a_\mu$ including the $\mu-t$
leptoquarks.  Although this $\mu-t$ leptoquark could imply a very large 
contribution to $a_\mu$ 
because of the large top quark mass, it could, however, 
give rise to flavor-changing processes such as $t \to c \gamma, c \mu^+ \mu^-$.
We do not consider this option.
Besides, we also have some sign differences in the main result.

The interaction Lagrangians for the $F=0$ and $F=-2$ 
($F$ is the fermion number) scalar leptoquarks are \cite{buch}
\begin{eqnarray}
\label{9}
{\cal L}_{F=0} &=& \lambda_L \overline{\ell_L} u_R {\cal S}_{1/2}^L
+ \lambda_R^* \overline{q_L} e_R (i \tau_2 {\cal S}^{R*}_{1/2} )
+ \tilde{\lambda}_L \overline{\ell_L} d_R \tilde{{\cal S}}_{1/2}^L + h.c. \;,\\
\label{10}
{\cal L}_{F=-2} &=& g_L \overline{q_L^{(c)}} i \tau_2 \ell_L {\cal S}_0^L
+ g_R \overline{u_R^{(c)}} e_R {\cal S}_0^R
+ \tilde{g}_R \overline{d_R^{(c)}} e_R \tilde{{\cal S}}_0^R
+ g_{3L}\overline{q_L^{(c)}} i \tau_2 \vec{\tau} \ell_L \cdot \vec{\cal S}_1^L
+ h.c.
\end{eqnarray}
where $q_L,\ell_L$ denote the left-handed quark and lepton doublets, 
$u_R,d_R,e_R$ denote the right-handed up-type quark, down-type quark, and 
lepton singlet, and $q_L^{(c)}, u_R^{(c)}, d_R^{(c)}$ denote the 
charge-conjugated fields.
The subscript on leptoquark fields denotes the weak-isospin of the leptoquark,
while the superscript ($L,R$) denotes the handedness of the lepton that
the leptoquark couples to.  The color indices of the quarks and leptoquarks
are suppressed.
The components of the $F=0$ leptoquark fields are
\begin{equation}
{\cal S}_{1/2}^{L,R} = \left ( \begin{array}{c}
        {S_{1/2}^{L,R} }^{(-2/3)} \\
        {S_{1/2}^{L,R} }^{(-5/3)}  \end{array} \right ) \;, \;\;\;\;\;
\tilde{{\cal S}}_{1/2}^L = \left( \begin{array}{c}
       \tilde{S}_{1/2}^{L(1/3)} \\
     - \tilde{S}_{1/2}^{L(-2/3)} \end{array} \right ) \;,
\end{equation} 
where the electric charge of the component fields is given in the 
parentheses, and the corresponding hypercharges are $Y({\cal S}_{1/2}^L)=
Y({\cal S}_{1/2}^R)=-7/3$ and $Y(\tilde{{\cal S}}_{1/2}^L)=-1/3$.
The $F=-2$ leptoquarks ${\cal S}_0^L, {\cal S}_0^R, \tilde{{\cal S}}_0^R$
are isospin singlets with hypercharges $2/3, 2/3, 8/3$, respectively, while
${\cal S}_1^L$ is a triplet with hypercharge $2/3$:
\begin{equation}
{\cal S}_1^L = \left( \begin{array}{l}
               { S_1^L }^{(4/3)} \\
               { S_1^L }^{(1/3)} \\
               { S_1^L }^{(-2/3)} \end{array} \right ) \;.
\end{equation}
The SU(2)$_L\times$ U(1)$_Y$ symmetry is assumed in the Lagrangians
of Eqs. (\ref{9}) and (\ref{10}).

To calculate the contribution to $a_\mu$ we start with the $F=0$ leptoquark
${\cal S}^{L,R}_{1/2}$ that has both the left- and right-handed couplings.
The other leptoquarks with either left- or right-handed
couplings are simply special cases of it. The Lagrangian can be rewritten as
\begin{equation}
\label{s12}
{\cal L}_{{\cal S}_{1/2}} = \bar \mu ( \lambda_L P_R + \lambda_R P_L ) c \, 
{\cal S}_{1/2}^{(-5/3)} \;\; + h.c. \;,
\end{equation}
where $P_{L,R}=(1\mp \gamma^5)/2$ and 
we explicitly write the second generation particles $\mu$ and $c$-quark.
The result can be easily obtained by some modifications on a $\mu\to e \gamma$
\cite{ott} calculation, as follows ($a_\mu$ is defined by ${\cal L}=(e/4 m_\mu)
a_\mu \bar \mu \sigma_{\alpha\beta} \mu F^{\alpha\beta}$)
\begin{equation}
\label{lq}
\Delta a_\mu( {\cal S}_{1/2} ) = - \frac{N_c}{16 \pi^2} 
\frac{m_\mu^2}{M^2_{ {\cal S}_{1/2}} } \left \{
( |\lambda_L|^2 + |\lambda_R|^2 ) ( Q_c F_5(x) - Q_S F_2(x) )
+ \frac{m_c}{m_\mu} {\cal R}e
(\lambda_L \lambda_R^*) ( Q_c F_6(x) - Q_S F_3(x) )
\right \} \;,
\end{equation}
where
\begin{eqnarray}
F_2(x) &=& \frac{1}{6 \, (1-x)^4} \, (1-6 \, x+3 \, x^2+2 \, x^3-6 \, x^2 \, \ln x) \; ,
\nonumber \\
F_3(x) &=& \frac{1}{(1-x)^3} \, (1-x^2+2 \, x \,\ln x) \; ,
\nonumber \\
F_5(x) &=& 
\frac{1}{6 \,(1-x)^4} \, (2+3 \, x-6 \, x^2+x^3+6 \, x  \, \ln x) \;,
\nonumber \\
F_6(x) &=&
\frac{1}{(1-x)^3} \, (-3+4 \, x-x^2-2 \,  \ln x) \; .
\nonumber 
\end{eqnarray}
In the above expression, $N_c=3, Q_c=2/3, Q_S =-5/3$, and 
$x=m_c^2/M^2_{ {\cal S}_{1/2}}$, and we have neglected terms proportional 
to $m_\mu^2/M^2_{ {\cal S}_{1/2}}$ in the parenthesis.  
Our expression agrees with that in Ref. \cite{equ}.

For the $F=-2$ leptoquarks only ${\cal S}_0^{L,R}$ has both the left- and
right-handed couplings.  The Lagrangian can be rewritten as 
\begin{equation}
\label{s0}
{\cal L}_{S_0}= 
\bar \mu ( g_L^* P_R + g^*_R P_L ) \, c^{(c)} \, {{\cal S}^*_0}^{(-1/3)}
\;\; + h.c. \;.
\end{equation}
The contribution to $a_\mu$ can be obtained from Eq. (\ref{lq}) with 
the following substitutions
\begin{equation}
m_c \to - m_c \;, \qquad Q_c \to Q_{c^{(c)}} \;, \qquad \lambda_{L,R} \to 
g^*_{L,R} \;,
\end{equation}
where $Q_{c^{(c)}}=-2/3$ and $Q_S = -1/3$ for this leptoquark.  

We note that our expression for $F=-2$ leptoquark agrees with Ref. \cite{ccc},
but we have a different expression for $F=0$ leptoquark.  Ref. \cite{ccc} does
not distinguish between these two types of leptoquarks.

\begin{figure}[bh]
\centering
\includegraphics[width=6in]{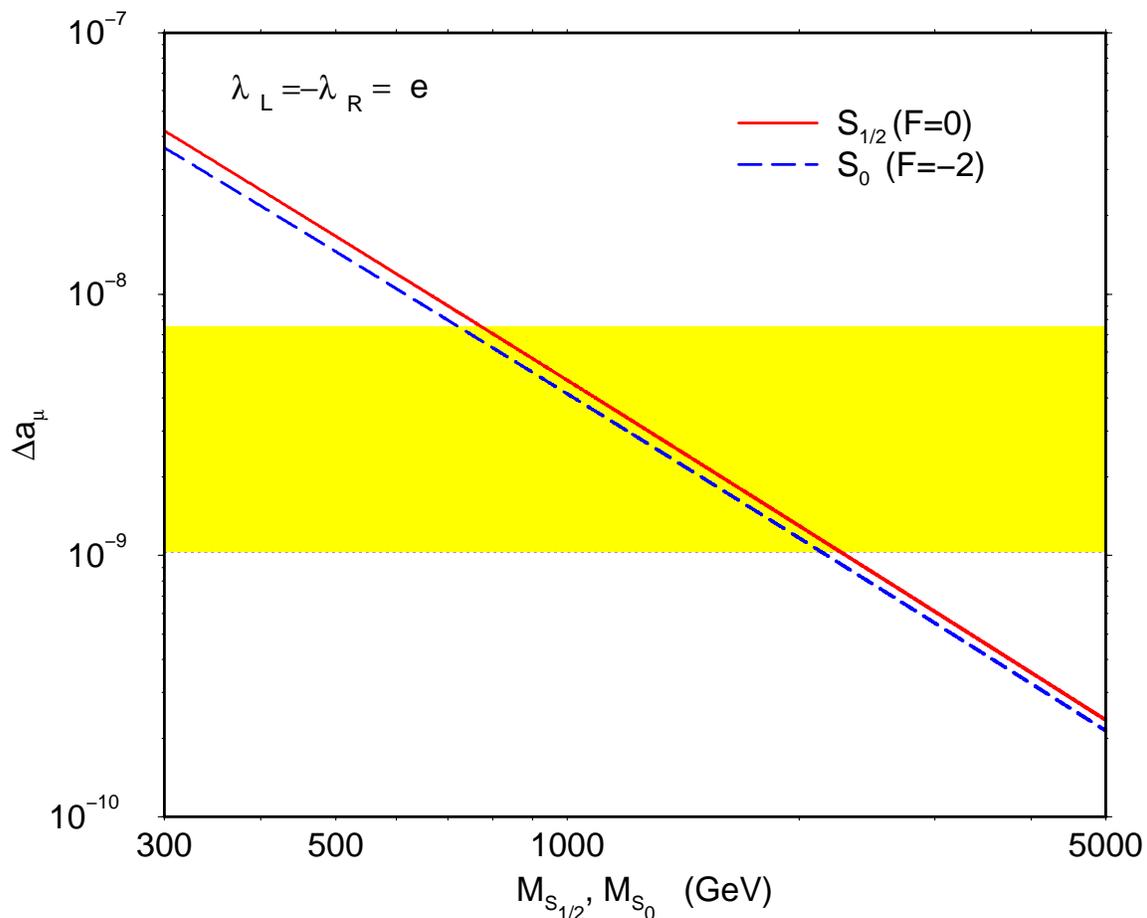}
\caption{\label{fig1}
Contributions to $\Delta a_\mu$ from the $F=0$ leptoquark ${\cal S}_{1/2}$
and the $F=-2$ leptoquark ${\cal S}_0$.  The shaded region is the 95\% C.L.
range of $\Delta a_\mu$ given in Eq. (\ref{95}).
}
\end{figure}

Next, we use our expressions to fit to $\Delta a_\mu$.  The range of
$\Delta a_\mu$ at 95\% C.L. ($\pm 1.96\sigma$) is
\begin{equation}
\label{95}
10.3\times 10^{-10} < \Delta a_\mu < 74.9 \times 10^{-10} \;.
\end{equation}
A rough estimate for the allowed range of $M_{\rm LQ}$ can be obtained by 
realizing the dominant term in Eq. (\ref{lq}).
In Eq. (\ref{lq}), the term with ${\cal R}e(\lambda_L \lambda_R^*)$ dominates
over the term with $(|\lambda_L|^2+|\lambda_R|^2)$, because of the
enhancement factor of $m_c/m_\mu$.  This is valid as long as $\lambda_L 
\approx \lambda_R$.  Also, the function $F_6(x) \to (-3-2 \ln x)$ and 
$F_3(x) \to 1$ when $x\to 0$.  Therefore, 
\begin{equation}
\Delta a_\mu ({\cal S}_{1/2}) \simeq \frac{-1}{8\pi^2} \, 
\frac{m_c m_\mu}{M^2_{{\cal S}_{1/2}}} \, {\cal R}e(\lambda_L \lambda_R^*) \;
(26) \;,
\end{equation}
where the numerical factor of 26 is estimated by varying $M_{{\cal S}_{1/2}}$
between $0.5-1.5$ TeV.
With the 95\% C.L. bound on $\Delta a_\mu$ we obtain
\begin{equation}
2.6\;{\rm TeV} < \frac{M_{{\cal S}_{1/2}}}
{\sqrt{- {\cal R}e(\lambda_L \lambda_R^*)}} < 7.2 \; {\rm TeV} \;.
\end{equation}
Similarly, for the $F=-2$ leptoquark ${\cal S}_0$ we obtain
\begin{equation}
2.5\;{\rm TeV} < \frac{M_{{\cal S}_0}}
{\sqrt{- {\cal R}e(g_L^* g_R)}} < 6.7 \; {\rm TeV} \;.
\end{equation}
If $\lambda_L=-\lambda_R=e$ and $g_L=-g_R=e$, where 
$e=\sqrt{4\pi \alpha_{\rm em}}$,
\begin{equation}
\label{est}
0.8\;{\rm TeV} < M_{{\cal S}_{1/2}} < 2.2 \; {\rm TeV} \qquad {\rm and} 
\qquad 
0.7\;{\rm TeV} < M_{{\cal S}_0} < 2.0 \; {\rm TeV} \;.
\end{equation}
We show in Fig. \ref{fig1} the contributions to $\Delta a_\mu$ from the 
$F=0$ and $F=-2$ leptoquarks ${\cal S}_{1/2}$ and ${\cal S}_0$ respectively,
using the exact expression of Eq. (\ref{lq}).  We have used $\lambda_L (g_L)=-
\lambda_R (g_R)=e$.  The shaded region is the 95\% C.L. range allowed as in
Eq. (\ref{95}).  One can see from the graph that the bounds on 
$M_{{\cal S}_{1/2}}$ and $M_{{\cal S}_0}$ are very close to the estimate in 
Eq. (\ref{est}).

What about the other leptoquarks that have only the left- or right-handed
coupling? We can use Eq. (\ref{lq}) with only $\lambda_L$ or $\lambda_R$, then
$\Delta a_\mu$ is given by
\begin{equation}
\Delta a_\mu  = - \frac{N_c}{16 \pi^2} 
\frac{m_\mu^2}{M^2_{\rm LQ} }
|\lambda_L|^2  ( Q_c F_5(x) - Q_S F_2(x) ) \;.
\end{equation}
The factor in the parenthesis is only a fraction of unity.  Thus, this
$\Delta a_\mu$ is suppressed by about $10^{-3}$ relative to the contributions
from ${\cal S}_{1/2}$ or ${\cal S}_0$. 
 Hence, the mass limits are weakened by a factor of
$\sqrt{10^{-3}}\approx 0.03$, which means the leptoquarks are to be lighter
than 100 GeV in order to explain the $a_\mu^{\rm exp}$.  It is obviously
ruled out by the Tevatron direct search limit on the second-generation
leptoquarks \cite{teva-lq} (see below).

We  note that these two leptoquarks also give rise to an electric dipole moment
(EDM) of muon, provided that ${\cal I}m(\lambda_L \lambda_R^*)$ is nonzero.
The contribution to EDM is given by
\begin{equation}
d_\mu =  \frac{e N_c}{32 \pi^2} 
\frac{m_c}{M^2_{\rm LQ}} \;
{\cal I}m (\lambda_L \lambda_R^*) ( Q_c F_6(x) - Q_S F_3(x) ) \;,
\end{equation}
where $d_f$ is defined by ${\cal L}=(-i/2) d_f \bar f \sigma_{\mu\nu} 
\gamma_5 f F^{\mu\nu}$.
Note that the same large numerical factor, scaling as 
$\ln(M^2_{\rm LQ}/m_c^2)$, is in the parenthesis.

We also note that the self-energy diagram of the muon 
with the leptoquark and charm quark inside
the loop gives a radiative correction to the muon mass.  We calculated this 
diagram and found that it has an UV divergent piece and a finite piece.  While
the divergent piece is absorbed into the renormalization constant, 
the finite piece
is given by $\delta m_\mu \sim (N_c \lambda^2/16\pi^2) m_c 
\ln(M_{\rm LQ}^2/m_\mu^2 )$.  Numerically, $\delta m_\mu$ 
is less than the observed muon mass for $\lambda \simeq e$ and 
$M_{\rm LQ}\simeq 1-2$ TeV,  such that $\delta m_\mu$
can be included into the definition of the pole mass without any fine tuning
problem, which gives the observed muon mass.

Summarizing, only the leptoquarks ${\cal S}_{1/2}$ and
${\cal S}_{0}$ that couple to both left- and right-handed muon can explain the
data on $\Delta a_\mu$, while the other leptoquarks alone cannot explain the
data.
In fact, it is advantageous to have the coexistence of other leptoquarks 
because they can satisfy constraints from other experiments and at the 
same time would not give any sizable contribution to $a_\mu$.


The most obvious limits on leptoquarks are the direct search limits at the
Tevatron $p\bar p$ collision and at the HERA $e^\pm p$ collision, based on
two NLO calculations \cite{nlo}.  Both
CDF and D\O\ searched for the first and second generation leptoquarks.  
Their limits are independent of the leptoquark couplings 
because the production is via the strong interaction.  The lower limits
on the first (LQ1) and second (LQ2) generation scalar leptoquarks are given by
\cite{teva-lq}
\begin{eqnarray}
M_{\rm LQ1} & > & 242 \; {\rm GeV} \;\;  {\rm for}\;\beta=1 \qquad
 \mbox{(CDF and D\O\ combined)} \;,     \nonumber\\
M_{\rm LQ2} & > & 202\;(160) \; {\rm GeV} \;\; {\rm for}\;\beta=1(0.5)\qquad
  \mbox{(CDF)} \;, \nonumber\\
M_{\rm LQ2} & > & 200 (180) \; {\rm GeV} \;\;  {\rm for}\;\beta=1(0.5)\qquad
  \mbox{(D\O)} \;,
\end{eqnarray}
where $\beta=B({\rm LQ} \to \ell q)$.
At HERA, the direct searches are limited to the first generation leptoquarks
and depend on the leptoquark couplings.  The best limits with
$\lambda=e$ are \cite{hera-lq}
\begin{eqnarray}
M_{\rm LQ1} & > & 280 \; {\rm GeV} \qquad \mbox{(ZEUS)} \;,\\
M_{\rm LQ1} & > & 275 \; {\rm GeV} \qquad \mbox{(H1)} \;.
\end{eqnarray}
The leptoquark solutions in Eq. (\ref{est}) are safe with these
limits.

There are also other existing constraints. Especially, if the 
first-second-generation universality is assumed for the leptoquarks, very
strong constraints come from low energy and high energy experiments
 \cite{bc,ours}.
Among the constraints the APV and the 
CC universality are the most relevant to leptoquarks.

\noindent
\underline{First-second-generation universality}

It is convenient to parameterize the effective interactions of leptoquarks
in terms of contact parameters $\eta^{\ell q}_{\alpha\beta}$,
where $\alpha$ and $\beta$ denote the chirality of the lepton and the quark,
respectively, when the mass of the leptoquarks are larger than
the energy scale of the experiment.  The contact parameters are defined by
\begin{equation}
{\cal L}_\Lambda = \sum_{\ell, q} \left \{
 \eta_{LL}^{\ell q} \overline{\ell_L} \gamma_\mu \ell_L 
\overline{q_L} \gamma^\mu q_L 
+\eta_{LR}^{\ell q} \overline{\ell_L} \gamma_\mu \ell_L 
\overline{q_R} \gamma^\mu q_R 
+\eta_{RL}^{\ell q} \overline{\ell_R} \gamma_\mu \ell_R 
\overline{q_L} \gamma^\mu q_L 
+\eta_{RR}^{\ell q} \overline{\ell_R} \gamma_\mu \ell_R 
\overline{q_R} \gamma^\mu q_R 
\right \} \;.
\end{equation}

The APV is measured in terms of weak charge $Q_W$.  The updated 
data  with an improved atomic calculation \cite{apv,atomic} 
is about $1.0 \sigma$ larger than the SM prediction, namely,
$\Delta Q_W \equiv Q_W({\rm Cs}) - Q_W^{\rm SM}({\rm Cs}) 
= 0.44 \pm 0.44$.
The contribution to $\Delta Q_W$ from the contact parameters 
is given by \cite{bc,ours}
\begin{equation}
\label{QW}
\Delta Q_W = ( -11.4\; {\rm TeV}^{2} ) \left[
-\eta_{LL}^{eu} + \eta_{RR}^{eu} - \eta_{LR}^{eu} + \eta_{RL}^{eu} \right ]
+ 
( -12.8\; {\rm TeV}^{2} ) \left[
-\eta_{LL}^{ed} + \eta_{RR}^{ed} - \eta_{LR}^{ed} + \eta_{RL}^{ed} \right ]
\;.
\end{equation}
Another important constraint is the CC universality.  It is  
expressed as $\eta_{CC} = \eta^{ed}_{LL} - \eta^{eu}_{LL} = (0.051 \pm 0.037)
\; {\rm TeV}^{-2}$.  These $\Delta Q_W$ and $\eta_{CC}$ are the two most
important constraints relevant to leptoquarks.  
With the first-second-generation universality
$\eta_{\alpha\beta}^{eu}=\eta_{\alpha\beta}^{\mu c}$ and 
$\eta_{\alpha\beta}^{ed}=\eta_{\alpha\beta}^{\mu s}$.
We are going to analyze the
leptoquark solutions that we found above with respect to these
two constraints.  Other high energy experiments such as HERA deep-inelastic
scattering, Drelly-Yan production, and LEPII hadronic cross sections also
constrained leptoquarks, but are relatively easy to satisfy with TeV mass
leptoquarks \cite{bc}.

For the $F=0$ leptoquark ${\cal S}_{1/2}$ with the interaction given in 
Eq. (\ref{s12}), the contributions to $\eta$ are
\begin{equation}
\eta^{\mu c}_{LR} = - \frac{|\lambda_L|^2}{2M^2_{{\cal S}_{1/2}}}\;, \qquad
\eta^{\mu c}_{RL} = - \frac{|\lambda_R|^2}{2M^2_{{\cal S}_{1/2}}}\;, 
\end{equation}
which are equal to $-(0.01 - 0.07) \; {\rm TeV}^{-2}$ for 
$\lambda_L=-\lambda_R=e$ and the mass range in Eq. (\ref{est}).
Similarly for the $F=-2$ leptoquark ${\cal S}_{0}$ with the interaction 
given in Eq. (\ref{s0}), the contributions to $\eta$ are
\begin{equation}
\eta^{\mu c}_{LL} = \frac{|g_L|^2}{2 M^2_{{\cal S}_{0}}}\;, \qquad
\eta^{\mu c}_{RR} = \frac{|g_R|^2}{2 M^2_{{\cal S}_{0}}}\;,
\end{equation}
which are equal to $0.01 - 0.08 \; {\rm TeV}^{-2}$ for $g_L=-g_R=e$ 
and the mass range in Eq. (\ref{est}).

Both of these leptoquarks do not contribute to $\Delta Q_W$ as the 
contributions get canceled.  While ${\cal S}_{1/2}$ does not contribute to
$\eta_{CC}$, ${\cal S}_{0}$ contributes to $\eta_{CC}$ but in the opposite
direction.  The lower mass range of ${\cal S}_{0}$ is then ruled out by 
the $\eta_{CC}$ constraint.

As mentioned above, coexistence of other leptoquarks
could satisfy the constraints on $\Delta Q_W$ and $\eta_{CC}$.
The $\Delta Q_W$ constraint can be satisfied
by the coexistence of either ${{\cal S}^R_{1/2}}^{(-2/3)}$ with interactions
$-\lambda_R\, \overline{e_R} \,d_L \, {{\cal S}^R_{1/2}}^{(-2/3)}+h.c.$, or
${\vec{\cal S}^L_1}$ with interactions
$-g_{3L} (\overline{u_L^{(c)}}e_L \, {\cal S}_1^{L(1/3)} +
\sqrt{2}\; \overline{d_L^{(c)}} e_L \; {\cal S}_1^{L(4/3)} ) +h.c.$ \cite{bc}.
The mass required to fit to $\Delta Q_W$ is
$M_{{\cal S}^R_{1/2}}=1.2$ TeV or $M_{\vec{\cal S}^L_1}=2.0$ TeV  
with electromagnetic coupling strength.  For such heavy leptoquarks
with only a left-handed or right-handed coupling, their 
contributions to $\Delta a_\mu$ are certainly negligible.  At the same time 
${\vec{\cal S}^L_1}$ contributes to $\eta_{CC}$ in the right direction, while
${{\cal S}^R_{1/2}}^{(-2/3)}$ does not.  

Summarizing, we can have the following three viable 
combinations of leptoquarks. 
\begin{enumerate}
\item ${{\cal S}_{1/2}}^{(-5/3)}$ and $\vec{\cal S}^L_1$.  The former
explains $\Delta a_\mu$ and the latter satisfies $\Delta Q_W$ and in the
right direction as $\eta_{CC}$.  This is the best scenario.

\item ${{\cal S}_{1/2}}^{(-5/3)}$ 
and ${{\cal S}^R_{1/2}}^{(-2/3)}$.  The former
explains $\Delta a_\mu$ and the latter satisfies $\Delta Q_W$.  They both
have no effect on $\eta_{CC}$, but it is fine.

\item${\cal S}_0$ and $\vec{\cal S}^L_1$.  The former
explains $\Delta a_\mu$ but violates $\eta_{CC}$. 
The latter can help pulling the leptoquark solution
within a reasonable deviation in $\eta_{CC}$ and still partially explaining
$\Delta Q_W$.
\end{enumerate}

\noindent
\underline{No first-second-generation universality}

In this case, virtually no constraints exist on the second generation 
leptoquarks.  The constraint of $D_s^+ \to \mu^+ \nu$ mentioned in 
Ref. \cite{ccc} only applies to a very low leptoquark mass, which has
already been ruled out by direct search \cite{teva-lq}.
There was a low-energy muon deep-inelastic scattering experiment on carbon
\cite{cern}.  An analysis \cite{cho} showed that this $\mu C$ experiment
results in a constraint
\begin{eqnarray}
2 \Delta C_{3u} - \Delta C_{3d}  &=& -1.505 \pm 4.92 \\
2 \Delta C_{2u} - \Delta C_{2d}  &=& 1.74 \pm 6.31
\end{eqnarray}
where $\Delta C_{2q}=(\eta_{LL}^{\ell q} - \eta_{LR}^{\ell q}
 + \eta_{RL}^{\ell q} - \eta_{RR}^{\ell q} )/(2\sqrt{2} G_F )$
and $\Delta C_{3q}=(-\eta_{LL}^{\ell q} + \eta_{LR}^{\ell q}
 + \eta_{RL}^{\ell q} - \eta_{RR}^{\ell q} )/(2\sqrt{2} G_F )$.
The leptoquark solutions of ${\cal S}_{1/2}$ and ${\cal S}_0$ give $\Delta 
C_{2q} =0$ and $\Delta C_{3q} \sim - 10^{-3}$.  Therefore, the constraint
from the $\mu C$ scattering is too weak to affect the leptoquark solutions.

We conclude that 
the $2.6\sigma$ deviation in the recent $a_\mu$ measurement 
places useful constraints on leptoquark models.  
To account for the $a_\mu$ data the leptoquark must have both
the left- and right-handed couplings to the muon.
Assuming that the couplings have electromagnetic strength, the
mass is restricted to be about $0.7 \; {\rm TeV} < M_{\rm LQ} < 2.2$ TeV.
If no first-second-generation universality is assumed, this mass range is
well above  the direct search limit at the Tevatron.
On the hand, if the first-second-generation universality is assumed, 
constraints also come from other low energy and high energy experiments,
among which the atomic-parity violation and charged-current universality are
the most important.  We have shown that coexistence with other leptoquarks
can satisfy these additional constraints and at the same time do not
affect the $a_\mu$.
Leptoquarks in such a mass range should be produced at the LHC via the
strong interaction.

I would like to thank Otto Kong for useful discussions, Paul Langacker 
for a correspondence, and special thanks to Stephan Narison for discussions
on hadronic uncertainties and renormalization.
This research was supported in part by the National Center for Theoretical
Science under a grant from the National Science Council of Taiwan R.O.C.


\begin{thebibliography}{99}

\bibitem{teva-lq}
CDF and D\O\ Collaborations (Carla Grosso-Pilcher {\it et al.}),
hep-ex/9810015;
CDF Coll., Phys. Rev. Lett. {\bf 79}, 4327 (1997);  
{\it ibid.} {\bf 81}, 4806 (1998);  
D\O\ Coll., Phys. Rev. Lett. {\bf 79}, 4321 (1997); 
{\it ibid.} {\bf 84}, 2088 (2000).

\bibitem{hera-lq}
H1 Coll., Eur. Phys. J. {\bf C11}, 447 (1999); 
{\it ibid.} {\bf C14}, 553 (2000);
ZEUS Coll., Eur. Phys. J. {\bf C16}, 253 (2000).

\bibitem{lq2}I. Bigi, G. Kopp,  and P. Zerwas, Phys. Lett. {\bf B166}, 
238 (1986);
G. Couture and H. K\"onig, Phys. Rev. {\bf D53}, 555 (1990);
S. Davidson, D. Bailey, and B. Campbell, Z. Phys. {\bf C61}, 613 (1994).

\bibitem{equ}
A. Djouadi, T. Kohler, M. Spira, and J. Tutas, Z. Phys. {\bf C46}, 679 (1990);

\bibitem{E821}H.N. Brown {\it et al.}, hep-ex/0102017.

\bibitem{am}A. Czarnecki and W. Marciano, hep-ph/0102122.

\bibitem{susy}L. Everett, G. Kane, S. Rigolin, and L. Wang, hep-ph/0102145;
J. Feng and K. Matchev, hep-ph/0102146;
E. Baltz and P. Gondolo, hep-ph/0102147;
U. Chattopadhyay and P. Nath, hep-ph/0102157;
S. Komine, T. Moroi, and M. Yamaguchi, hep-ph/0102204.

\bibitem{Z}D. Choudhury, B. Mukhopadhyaya, and S. Rakshit, hep-ph/0102199;
T. Huang, Z. Lin, L. Shan, and X. Zhang, hep-ph/0102193.

\bibitem{ccc}D. Chakraverty, D. Choudhury, and A. Datta, hep-ph/0102180.

\bibitem{lq} U. Mahanta, hep-ph/0102176, hep-ph/0102211.

\bibitem{preon}K. Lane, hep-ph/0102131.

\bibitem{buch}
W. Buchm\"{u}ller, R. R\"{u}ckl, and D. Wyler, Phys. Lett. {\bf B191}, 
442 (1987);
J. Hewett and T. Rizzo, Phys. Rev. {\bf D56}, 5709 (1997). 

\bibitem{ott} K. Cheung and O. Kong, hep-ph/0101347.


\bibitem{nlo}
M. Kramer, T. Plehn, M. Spira, and P. Zerwas, Phys. Rev. Lett. {\bf 79},
341 (1997); 
T. Plehn, H. Spiesberger, M. Spira, and P. Zerwas, Z. Phys. {\bf C74}, 611
(1997).

\bibitem{bc}V. Barger and K. Cheung, Phys. Lett. {\bf B480}, 149-154 (2000).

\bibitem{ours}V. Barger, K. Cheung, K. Hagiwara, and D. Zeppenfeld,
Phys. Rev. {\bf D57}, 391 (1998); 
K. Cheung, hep-ph/9807483; D. Zeppenfeld and K. Cheung, hep-ph/9810277.

\bibitem{apv}
S.C. Bennett and C.E. Wieman, Phys. Rev. Lett. {\bf 82}, 2484 (1999);
C.S. Wood, {\it et al.}, Science {\bf 275}, 1759 (1997).

\bibitem{atomic}A. Derevianlo, Phys. Rev. Lett. {\bf 85}, 1618 (2000);
P. Langacker, hep-ph/0102085.

\bibitem{cern}A. Argento {\it et al.}, Phys. Lett. {\bf B120}, 245 (1983).

\bibitem{cho}
G. Cho, K. Hagiwara, and S. Matsumoto, Eur. Phys. J. {\bf C5}, 155 (1998).

\end{thebibliography}
\end{document}